\documentstyle[amsmath,amssymb,graphicx,multicol]{article}

\def\be{\begin{eqnarray}}
\def\ee{\end{eqnarray}}
\def\nn{\nonumber}

\def\tr{{\rm tr}\,}

\def\HL{Hall-Littlewood\ }
\def\l[{\phantom.[}

\hoffset= - 1.0in         

\title{{\bf Torus HOMFLY as the Hall-Littlewood Polynomials  } \vspace{.2cm}}
\author{{\bf A.Mironov}\footnote{ {\small {\it
Lebedev Physics Institute} and {\it ITEP, Moscow, Russia}};
mironov@itep.ru; mironov@lpi.ru}, {\bf A.Morozov}\thanks{{\small
{\it ITEP, Moscow, Russia}}; morozov@itep.ru}, {\bf
Sh.Shakirov}\thanks{{\small {\it Department of Mathematics,
University of California, Berkeley, USA}, {\it Center for
Theoretical Physics, University of California, Berkeley,
USA}}}\date{ }}

\begin{document}
 \maketitle

\vspace{-5.5cm}

\begin{center}
\hfill FIAN/TD-02/12\\
\hfill ITEP/TH-08/12\\
\end{center}

\vspace{5cm}

\centerline{ABSTRACT}

\bigskip

{\footnotesize We show that the HOMFLY polynomials for torus knots
$T[m,n]$ in all fundamental representations are equal to the
Hall-Littlewood polynomials in representation which depends on $m$,
and with quantum parameter, which depends on $n$. This makes the
long-anticipated interpretation of Wilson averages in $3d$
Chern-Simons theory as characters precise, at least for the torus
knots, and calls for further studies in this direction. This fact is
deeply related to Hall-Littlewood-MacDonald duality of character
expansion of superpolynomials found in \cite{MMSS}. In fact, the
relation continues to hold for {\it extended} polynomials, but the
symmetry between $m$ and $n$ is broken, then $m$ is the number of
strands in the braid. Besides the HOMFLY case with $q=t$, the torus
{\it super}polynomials are reduced to the single Hall-Littlewood
characters in the two other distinguished cases: $q=0$ and $t=0$.}

\vspace{1.cm}

\section{Introduction}

$3d$ Chern-Simons theory (CST) \cite{CS} is characterized by the next level of complexity
as compared with the $2d$ conformal field theory (CFT) \cite{CFT}.
After a breakthrough in CFT
induced by the recent study of AGT relations \cite{AGT},
the time is coming for similar studies in CST.
Like CFT \cite{CFTfreef}, CST is essentially a Gaussian (free field) theory,
which is obvious in special gauges, $A_0=0$ \cite{MoSm} or $\bar A=0$ \cite{DMS1}.
Moreover, at the level of unknots, CST basically reduces to CFT \cite{CS}.
A non-trivial addition to CFT at the level of CST is a new set of observables:
Wilson averages along {\it arbitrary} (linked) contours in $3d$ space (knots and links
\cite{CSknot}),
with a non-trivial projection to $2d$ space,
called HOMFLY polynomials \cite{HOMFLY}:
\be
\left.H^{{\rm link}}_R(t|A)\right|_{A=t^N} =
\ \left< \rm Tr_R \exp \oint_{{\rm link}} {\cal A} \right>
\ee
where the CST average at the r.h.s. is defined with the help
of the CST action
$\frac{\kappa}{4\pi}\int_{d^3x} \tr\left({\cal A}d{\cal A} + \frac{2}{3}{\cal A}^3\right)$,
parameter $t = \exp\left(\frac{2\pi i}{\kappa+N}\right)\equiv
\exp\left(\frac{2\pi i}{\tilde\kappa}\right)$
and $R$ is a representation of the gauge group $SU(N)$, labeled by a Young diagram
$R = \{r_1\geq r_2\geq \ldots \geq 0\} = [r_1,r_2,\ldots] = [\ldots,\underbrace{2,\ldots,2}_{m_2},
\underbrace{1,\ldots,1}_{m_1}]$.
We denote the number of non-vanishing $r_i$ in $R$ (the number of lines in the Young diagram)
through $l(R)=\sum_j m_j(R)$, where $m_j(R)$ is the number of lines with length $j$
in the Young diagram $R$.

\bigskip

Remarkably, $H(t|A)$ is essentially (with a proper normalization) a
polynomial both in $t$ and $A$, and the rank of the gauge group
enters the answer only through $A$ (this property is obvious from
the alternative description of $H(t|A)$ through the skein relations
\cite{skein}). As a topological theory partition function, the CST
functional integral basically reduces to the $0d$ (matrix) model
\cite{CSmamo}
\be\label{CSmmp}
\int \prod_{i=1}^N da_i \exp
\left(\frac{\tilde\kappa\eta a_i^2}{4\pi i}\right) \prod_{i<j}
\sinh^2\frac{a_i-a_j}{2} =\ \Big<\Big< 1 \Big>\Big>_\eta
\ee
with
$\eta=1$, which makes it even closer to CFT similarly reduced from
$2d$ field theory to  "conformal" \cite{confmamo} or Dotsenko-Fateev
\cite{DFmamo} matrix models. Unfortunately, it is still unknown,
what should be inserted into the integrand to reproduce arbitrary
HOMFLY polynomials.

However, this {\it is} known in the special case of {\it torus} links $T[m,n]$ \cite{torus,CSmamo}:
the relevant expression is
\be
H^{m,n}_R(t|A) \sim \ \Big<\Big< \chi_{\phantom{}_{Y_{\eta}}}(e^a)\,\chi_R(e^{ma})
\Big>\Big>_{\eta=m/n}\
=\sum_{Q\vdash m|R|} C_R^Q\ \Big<\Big< \chi_{\phantom{}_{Y_{\eta}}}(e^a)\,\chi_Q(e^{a})
\Big>\Big>_{\eta=m/n}
\label{Hchichi}
\ee
where the coefficients $C_R^Q$ are determined from the decomposition
$\chi_R(e^{ma}) = \sum_{Q\vdash m|R|}
C_R^Q\, \chi_Q(e^{a})$ (describing Adams' operation \cite{torus}), and
$Y_{\eta}$ is the Young diagram with the line lengths $Y_i=(\eta-1)(N-i)$ so that
$\chi_{\phantom{}_{Y_{\eta}}}(e^a) =
\prod_{i<j} \frac{\sinh\eta(a_i-a_j)/2}{\sinh(a_i-a_j)/2}$, being
continued to non-integer $\eta=m/n$.
This puts CST into the standard context of the modern matrix model theory,
where the most important observables are averages of characters \cite{unit}.
In matrix models of the fundamental nature these averages are expected to be
of the type of Selberg integrals, i.e. are themselves the characters.
This is one of the intuitive lines, which lead one to a hypothesis that the
HOMFLY polynomials,
as well as their $\beta$-deformations \cite{beta}, the superpolynomials \cite{sups}-\cite{MMSS},
should be interpreted as characters.

Of course, the Wilson $P$-exponent is a kind of character from the very
beginning, but that of an infinite dimensional loop group. In fact, here we face
the well-known phenomenon that the correlators in quantum integrable systems are
associated with a classical integrable system (typically, their generation function is
a $\tau$-function of an integrable system) \cite{qclint}. The concept
$<{character}>\ = {character}$ is just of this type. Unfortunately, we are still far
from its understanding, thus, reduction to the matrix model level, where something is
already known, makes the hypothesis much better established and concrete.

A direct application of this idea to (\ref{Hchichi}) implies the use of
a Selberg-like formula $<<\chi_{\phantom{}_{Y_{\eta}}} \chi_Q>>_{\eta=m/n}\sim t^{\frac{n}{m}(\nu_Q-\nu_{Q'})}\chi_Q^*$, where
$\chi_Q^*$ is the value of character at the special point $e^{a_i}=t^{i-1}$. This results into the
$W$-representation \cite{Wrep} of {\it extended} HOMFLY polynomials \cite{torus,DMMSS}:
\be
H^{m,n}_R\{t|p\} = \sum_{Q\vdash m} t^{\frac{n}{m}(\nu_Q-\nu_{Q'})}C_R^Q \chi_Q\{p\}
= t^{\frac{n}{m}\hat W_0} \widehat{Ad}_m \chi_R\{p\}
\label{chi}
\ee
where $t$ and $n$ appear only in the first factor, i.e. in the combination $t^n$,
$\nu_Q = \sum_i (i-1)q_i$ for $Q=\{q_1\geq q_2\geq\ldots\geq 0\}$,
and the Adams operation acts on the time-variables:
\be
\widehat{\rm Ad}_m f\{p_k\} = f\{p_{mk}\}
\ee
This is a very useful formula, but it is, once again,
a character {\it decomposition}, not a character {\it interpretation} of
the HOMFLY polynomials.

In what follows it will be convenient to use a different normalization of
HOMFLY polynomials (and superpolynomials), such that the "highest" character
$\chi_{mR}$ enters the character expansion (\ref{chi}) with the unit coefficient.
We denote such "renormalized" polynomials by $\bar H$ (and $\bar P$).

\section{HOMFLY case}

In this letter we convert the above {\it hypothesis} into a precise {\it statement},
which is much stronger than (\ref{chi}):
at least, for torus links $T[m,n]$
and, at least, for the fundamental representations
$R = [1^s] = [\underbrace{1,\ldots,1}_s]$
\be
\boxed{
\phantom{5^{5^{5^5}}}
\bar H^{m,n}_{[\underbrace{1,\ldots,1}_s]}\{t|p\} =
t^{\frac{1}{2}ns(m-s)} H^{m,n}_{[\underbrace{1,\ldots,1}_s]}\{t|p\} =
L_{[\underbrace{m,\ldots,m}_s]}\{t^n|p\}
\phantom{5^{5^{5^5}}}
}
\label{main_0}
\ee
or $H^{m,n}_{[1^s]}\{t|p\} = t^{\frac{n}{m}(\nu_M-\nu_{M'})} L_{M}\{t^n|p\}$
where $M\equiv  [m^s] = [\underbrace{m,\ldots,m}_s] $.

Here
$L_Q\{t|p\}$ are the \HL  polynomials, i.e. the polynomials produced from the
MacDonald polynomials $M_Q\{q,t|p\}$ at $q=0$.\footnote{
Likewise the Schur polynomials $\chi_Q\{p\}$ arise from $M_Q\{q,t|p\}$
when $q=t$,
the elementary symmetric polynomials $m_Q\{p\}$ when $q=1$
(in both cases the dependence of $t$ disappears), the
Jack polynomials $J_Q\{\beta|p\}$ are produced when $t=q^\beta$ and $q,t\rightarrow 1$.
Note that while the superpolynomials in (\ref{chisup}) below
reduce to the HOMFLY ones in (\ref{chi}) at $q=t$,
the same HOMFLY polynomials are expressed by (\ref{main_0})
through the \HL ones associated with $q=0$.
}
Relation (\ref{main_0}), like (\ref{chi}), is actually true at the level of
{\it extended} HOMFLY polynomials \cite{DMMSS,knMMM12},
which depend on the $m$-strand braid representation of the link/knot
and on infinitely many time variables $\{p_k\}$
instead of a single one $A$.
The topologically invariant HOMFLY polynomials themselves
emerge at the one-dimensional locus
\be
p_k=p_k^* = \frac{1-A^k}{1-t^k}
\label{tolo}
\ee

Actually this theorem (\ref{main_0}) is intimately related to
the study of character expansions of HOMFLY and superpolynomials
in \cite{DMMSS} and \cite{knMMM12} respectively,
and to the \HL-MacDonald duality of these expansions
discovered recently in \cite{MMSS}.
From this perspective, a somewhat better formulation of (\ref{main_0})
with $s=1$
is in terms of the generating function
(which is in the number $m$ of strands!)
\be
\boxed{
\sum_m z^m
\bar H^{m,n}_{[1]}\{t|p\} = \
\frac{1}{1-t^n}\ \exp \left(\ \sum_{k=1}^\infty \frac{1-t^{nk}}{k}p_k z^k\right)
}
\label{genf_0}
\ee
After that one can apply the Cauchy expansion formula
\be
\exp \left(\sum_{k=1}^\infty \frac{1-t^k}{k} p_k\bar p_k\right)
= \sum_Q \frac{L_Q\{t|p\} L_Q\{t|\bar p\}}{||L_Q||^2}
\label{HLCau}
\ee
with
\be
||L_Q||^{-2} = (1-t)^{l(Q)}\prod_j [m_j(Q)]_t!\equiv {(1-t)^{l(Q)}\over\omega_Q}
\ee
in different ways, by making different choices for $\bar p_k$. For instance:
\begin{itemize}
\item
For $\bar p_k = z^k$ eqs.(\ref{genf_0}) and (\ref{HLCau}) imply (\ref{main_0}),
since $L_Q\{t|\bar p_k=z^k\} = z^{|Q|}\delta_{l(Q),1}$,
where $l(Q)$ is the number of lines in the Young diagram $Q$.
Of course, the inverse claim is also true: (\ref{main_0}) and (\ref{HLCau}) imply (\ref{genf_0}).
\item
For $\bar p_k = \frac{1-t^{nk}}{1-t^k} z^k$
eqs.(\ref{genf_0}) and (\ref{HLCau}) imply instead
\be
\bar H^{m,n}_{[1]}\{t|p\}
= \sum_{\stackrel{Q\vdash m}{l(Q)\leq n}} h_Q L_Q\{t|p\}
\label{HLexpan_0}
\ee
where the expansion coefficients are themselves
essentially
the values of
characters $L_Q$ at a special point:\footnote{
It can be useful to keep in mind a more general formula for the
\HL polynomials after the Miwa transform $p_k = \sum_{i=1}^r x_i^k$:
$$
L_Q[t|x_1,\ldots,x_r] = {\rm Symm}\left(x_1^{Q_1}\ldots x_r^{Q_r}
\prod_{i<j} \frac{x_i-tx_j}{x_i-x_j}\right)
$$
}
\be
h_Q = \frac{L_Q\left\{t\Big|\bar p_k=\frac{1-t^{nk}}{1-t^k}\right\}}{(1-t^n)
||L_Q||^2}
= \frac{L_Q\left[t\Big|1,t,\ldots,t^{(n-1)}\right]}{(1-t^n)||L_Q||^2}
= t^{\nu_Q} (1-t)^{l(Q)-1} \prod_{j=1}^{l(Q)-1} [n-j]_t
\label{hco_0}
\ee
This reproduces the $h_Q$ coefficients obtained in \cite{MMSS} and further
generalized ($\beta$-deformed) there to the case of superpolynomials.
{\bf Note that unlike (\ref{main_0}), the quantum parameter
in the \HL polynomials in (\ref{HLexpan_0}) is $t$, not $t^n$.}
In (\ref{hco_0}) the time argument of the \HL polynomial
is actually the sum of $n$ terms:
$p_k=\frac{1-t^{nk}}{1-t^k} = 1 + t^k + t^{2k} + \ldots + t^{(n-1)k}$.
As usual with characters, this means that $L_Q$ with this argument
is non-vanishing only when the number of lines in the Young
diagram is no more than $n$: $l(Q)\leq n$, which is obvious at the r.h.s.
of (\ref{hco_0}) and very important for the superpolynomial generalizations
in \cite{MMSS}, where this selection rule remains preserved after the $\beta$-deformation.
\item
One can also use a more familiar version of the Cauchy formula,
\be
\exp \left(\sum_{k=1}^\infty \frac{p_k\bar p_k}{k} \right)
= \sum_Q \chi_Q\{p\}\chi_Q\{\bar p\}
\ee
to expand (\ref{main_0}) into the Schur characters $\chi_Q$, but the formulas
emerging in this way
are bilinear in $\chi\{p\}$ (and also bilinear in $\chi\{\bar p\}$
with certain choices of $\bar p_k$) and do not seem very interesting.
\item
The fact that the quantum parameter in (\ref{main_0})
is $t^n$ is in clear accordance with (\ref{chi}).
However, any direct relation between (\ref{chi}) or its underlying matrix model
integrals and (\ref{main_0}) remains obscure.
\end{itemize}

\section{On colored HOMFLY polynomials}

For $m=1$, i.e. for the unknot,  $H_R^{1,n}\{t|p\} = t^{\nu_{R}-\nu_{R'}}\chi_R\{p\}$
and this is proportional to $L_R\{t^n|p\}$ for $R = [1^s]$,
since $L_{[1^s]}\{t|p\} = \chi_{[1^s]}\{p\}$ does not depend on $t$.
Thus {\bf the extended HOMFLY polynomial  for the unknot {\it is} a character}
in any representation $R$.

\bigskip

However, in general for non-fundamental representations
$R \neq [1^s]$ the Hall-Littlewood decomposition
(\ref{main_0}) of the {\it colored} HOMFLY polynomials becomes more involved.
For example,
\be
H_{[2]}^{m,n}\{t|p\}
= t^{-n(2m-1)}L_{[2m]}\{t^{2n}|p\} + t^{-n(m-2)}L_{[m,m]}\{t^n|p\}=\nn \\
= t^{-n(2m-1)}\left(L_{[2m]}\{t^n|p\} + (1-t^n)\sum_{j=1}^{m-1} t^{nj} L_{[2m-j,j]}\{t^n|p\}
+ t^{nm} L_{[m,m]}\{t^n|p\}
\right)
\label{H2}
\ee
Thus for $R\neq [1^m]$ the \HL basis provides only a character {\it decomposition},
not {\it interpretation}.

\bigskip

The counterparts of the two versions of eq.(\ref{H2}) for other non-fundamental representations
remains to be found.
The second line is straightforwardly obtained from the \HL expansion (like (\ref{HLexpan_0})) of (\ref{chi}),
taken at $n=1$, by simply changing $t$ for $t^n$.
The first line, providing a more concise expression is not so straightforward
and it is obscure, whether it has any generalization at all.

Actually, the structure is simple at $n=0$, when the $t$-dependence disappears,
and $H_{[R]}^{m,0}\{p\} = \widehat{Ad}_m\chi_R\{p\}$:
\be
\nn\\
H_{[1]}^{m,0}\{p\} =    L_{[m]}\{1|p\}, \nn \\ \nn \\
H_{[2]}^{m,0}\{p\} =  L_{[2m]}\{1|p\} +  L_{[m,m]}\{1|p\}\nn\\
H_{[1,1]}^{m,0}\{p\} =   L_{[m,m]}\{1|p\}, \nn \\ \nn \\
H_{[3]}^{m,0}\{p\} =   L_{[3m]}\{1|p\} +   L_{[2m,m]}\{1|p\} +   L_{[m,m,m]}\{1|p\}\nn \\
H_{[2,1]}^{m,0}\{p\} =   L_{[2m,m]}\{1|p\} + 2   L_{[m,m,m]}\{1|p\} \nn \\
H_{[1,1,1]}^{m,0}\{p\} =    L_{[m,m,m]}\{1|p\} \nn \\ \nn \\
H_{[4]}^{m,0}\{p\} =  L_{[4m]}\{1|p\} + L_{[3m,m]}\{1|p\}
+ L_{[2m,2m]}\{1|p\} +  L_{[2m,m,m]}\{1|p\} + L_{[m,m,m,m]}\{1|p\}\nn \\
H_{[3,1]}^{m,0}\{p\} = L_{[3m,m]}\{1|p\} + L_{[2m,2m]}\{1|p\}
+ 2L_{[2m,m,m]}\{1|p\} + 3L_{[m,m,m,m]}\{1|p\}\nn \\
H_{[2,2]}^{m,0}\{p\} =  L_{[2m,2m]}\{1|p\} +  L_{[2m,m,m]}\{1|p\}
+ 2 L_{[m,m,m,m]}\{1|p\}\nn \\
H_{[2,1,1]}^{m,0}\{p\} =  L_{[2m,m,m]}\{1|p\} + 3 L_{[m,m,m,m]}\{1|p\} \nn \\
H_{[1,1,1,1]}^{m,0}\{p\} = L_{[m,m,m,m]}\{1|p\}, \nn \\ \nn \\
\ldots
\ee
or
\be
\begin{array}{c|ccccc}
& [4m] & [3m,m] & [2m,2m] & [2m,m,m] & [m,m,m,m]\\
\hline
\l[4] & 1 & 1 & 1 & 1 & 1 \\
\l[3,1] & &1 &1& 2 & 3 \\
\l[2,2] &&& 1 & 1 & 2 \\
\l[2,1,1] &&&& 1 & 3 \\
\l[1,1,1,1] &&&&&1
\end{array}
\ee
The next portion of coefficients is:
\be
\begin{array}{c|ccccccc}
& [5m] & [4m,m] & [3m,2m] & [3m,m,m] & [2m,2m,m] & [2m,m,m,m] & [m,m,m,m,m] \\
\hline
\l[5] &1&1&1&1&1&1&1 \\
\l[4,1] &&1&1&2&2&3&4\\
\l[3,2] &&&1&1&2&3&5\\
\l[3,1,1] &&&&1&1&3&6\\
\l[2,2,1] &&&&&1&2&5\\
\l[2,1,1,1] &&&&&&1&4\\
\l[1,1,1,1,1] &&&&&&&1\\
\end{array}
\ee

However, this nice triangular structure,
while preserved for all $n$ in the first line of (\ref{H2}),
gets destroyed in the second line, and for other $R > [2]$:
the coefficient matrices $\eta_{RQ}(t)$ in
\be
H_{R}^{m,n}\{t|p\} = t^{n(\nu_{mR}-\nu_{mR'})/m}\sum_{Q\vdash m|R|}
\eta_{RQ}(t^n)L_{Q}\{t^n|p\}
\label{HR}
\ee
are not triangular and even square matrices for $t\neq 1$.

For $R=2$ one has instead: $\eta_{[2],[2m]}(t) = 1,\ \eta_{[2],[2m-i,i]}(t) = t^i(1-t),\ 0<i<m,
\ \ \eta_{[2],[m,m]}(t)=t^m$.
Note that in (\ref{HR}) the argument of $\eta$ is $t^n$, not just $t$.

For the three-box diagrams the non-vanishing $\eta_{RQ}$'s are:
\be
R=[2,1]: \ \ \ \eta_{[2,1],[2m,m]} = 1,\ \eta_{[2,1],[2m-i,m,i]} = t^i(1-t),\ \ \eta_{[2,1],[m,m,m]}=t^m(1+t)
\ee
and
\be
 R=[3]: \ \ \ \eta_{[3],[3m]} = 1,\ \eta_{[3],[3m-i,i]} = t^i(1-t^2),
 \ \ \eta_{[3],[2m,m]}=t^m(1-t^2+t^{m+1}),\nn \\
 \eta_{[3],[3m-i-j,i,j]} = t^{i+2j}(1-t)(1-t^2),\ 0<i\leq j < m, \nn \\
  \eta_{[3],[2m-j,m,j]} = t^{m+2j}(1-t)(1-t+t^{m+1-j}),\ \
 \eta_{[3],[m,m,m]} = t^{3m}
\ee
In the last case one can make a step towards the first line of (\ref{H2}):
by switching from $L_{3m}\{t^n|p\}$ to $L_{3m}\{t^{3n}|p\}$ one obtains
a significant simplification:
\be
H_{[3]}^{m,n}\{t|p\} = t^{{3\over 2}n(3m-1)} \left( L_{[3m]}\{t^{3n}|p\}
+ t^{(2m+1)n} L_{[2m,m]}\{t^{n}|p\} + \sum_{j=1}^m t^{n(2m+1+j)}(1-t^n)L_{[2m-j,m,j]}\{t^n|p\}
+ \right.\nn \\ \left. \phantom{\sum_{j=1}^m}
+ t^{(3m+1)n}(1+t^n-t^{2n})L_{[m,m,m]}\{t^n|p\}\right)
\label{H3}
\ee

\bigskip

In fact, for any $s$, in addition to the fundamental representation $R=[1^s]$,
there is always one more combination of HOMFLY polynomials with
different $R$ of the same size $|R|=s$, which {\it is} a character.
To see this, let us introduce the operator
\be
\hat H_{m,n}  =  \widehat{Ev}_{n/m} \widehat{Ad}_m
\ee
where $\widehat{Ev}_k = e^{k\hat W_0}$.
According to (\ref{chi}) the HOMFLY polynomial
\be
H_R^{m,n} = \hat H_{m,n} \chi_R
\label{hatHchi}
\ee
In particular, (\ref{main_0}) states that
\be
\hat H_{m,n}
 \chi_{[1^s]}
 = t^{-\frac{1}{2}ns(m-s)} L_{[m^s]}(t^n)
\ee
It follows that when the same operator is applied to
a single time variable $p_k$, one gets:
\be
\hat H_{m,n}  p_k = \hat H_{m,n} \widehat{Ad}_k p_1
= \widehat{Ev}_{n/m} \widehat{Ad}_m \widehat{Ad}_k
\chi_{[1]} =
\widehat{Ev}_{nk/mk} \widehat{Ad}_{mk}  \chi_{[1]} =
\hat H_{nk,mk} \chi_{[1]} =
 t^{-\frac{1}{2}nk(mk-1)}
L_{mk}(t^{nk})
\ee
At the same time $p_k$ is a linear combination of
Schur functions of the weight $k$,
\be
p_k = \sum_{Y \in k} \varphi^Y_k \chi_{Y}\{p\}
\ee
and, together with (\ref{hatHchi}), this implies
that a certain linear combination of HOMFLY polynomials
is indeed a character $L_{mk}\{t^{nk}|p\}$.

In particular, for 3-box diagrams one has
\be
\bar H_{[3]} - \bar H_{[2,1]} + \bar H_{[1,1,1]}=
L_{[3m]}\{t^{3n}|p\}
\ee
what can be used to explain the simplification in (\ref{H3}).

\section{Superpolynomials at $q=0$ and $t=0$}

For generic $q\neq t$ the extended torus {\it super}polynomials
also possess character decompositions:
in MacDonald \cite{DMMSS} and Hall-Littlewood \cite{MMSS} polynomials:\footnote{
The first few MacDonald and \HL polynomials are:
$$
M_{[1]} = L_{[1]} = p_1, \ \ \
M_{[2]} = \frac{(1+t)(1-q)p_2 + (1-t)(1+q)p_1^2}{2(1-qt)}\ \stackrel{q=0}{\longrightarrow}\
L_{[2]} = \frac{(1+t)p_2 + (1-t)p_1^2}{2},
\ \ \ M_{[1,1]} = L_{[1,1]} = \frac{-p_2+p_1^2}{2},
$$ $$ \!\!\!\!\!\!\!
M_{[3]} = \frac{(1-q)(1-q^2)(1-t^3)}{3(1-t)(1-tq)(1-tq^2)}p_3
+ \frac{(1-t^2)(1-q^3)}{2(1-tq)(1-tq^2)}p_2p_1
+ \frac{(1+q)(1-q^3)(1-t)^2}{6(1-q)(1-tq)(1-tq^2)}p_1^3
\ \stackrel{q=0}{\longrightarrow}\
L_{[3]} = \frac{1+t+t^2}{3}p_3 + \frac{1-t^2}{2}p_2p_1 + \frac{(1-t)^2}{6}p_1^3,
$$ $$ \!\!\!\!\!\!\!
M_{[2,1]} = -\frac{(1-q)(1-t^3)}{3(1-t)(1-qt^2)}p_3 + \frac{(1+t)(t-q)}{2(1-qt^2)}p_2p_1
+ \frac{(1-t)(2+q+t+2qt)}{6(1-qt^2)}p_1^3
\ \stackrel{q=0}{\longrightarrow}\
L_{[2,1]} = -\frac{1+t+t^2}{3}p_3 + \frac{t(1+t)}{2}p_2p_1 + \frac{(1-t)(2+t)}{6}p_1^3,
$$ $$
M_{[1,1,1]} = L_{[1,1,1]} = \frac{1}{3}p_3 - \frac{1}{2}p_2p_1 + \frac{1}{6}p_1^3 \nn
$$
Re-expansion of MacDonald polynomials into the \HL and Schur functions
and vice versa is by itself a rather tedious calculation.
}
\be
\bar P^{m,n}_{[1]}\{q,t|p\} = q^{k\nu_{1^m}}
\sum_{Q\vdash m} t^{k\nu_Q}q^{-k\nu_{Q'}}c_Q M_Q\{p\}
\stackrel{k=0}{\longrightarrow}  \sum_{\stackrel {Q\vdash m}{l(Q)\leq r}} h_Q^{(r)}(q,t) L_Q\{t|p\}
\label{chisup}
\ee
with $n = mk+r$.
The coefficients $c_Q$ and $h_Q$ are rather sophisticated
and known in full generality only for $r=1,2,3,4$,
see \cite{DMMSS} and \cite{MMSS} for various examples.

However, this character decomposition drastically simplifies
and reduces to a single Hall-Littlewood character
not only in the HOMFLY case of $q=t$ in (\ref{main_0}),
but also in two other cases, $q=0$ and $t=0$:
\be
\boxed{
\bar P^{m,n}_{[1]}\{q=0,t|p\} =\ L_{[m]}\{t|p\}\phantom{\frac{1}{t}}\!\!\! }\ \  =
\frac{1}{1-t}\chi_{[m]}\big\{(1-t^k)p_k\big\},
\label{Pq0}\\
\boxed{
\bar P^{m,n}_{[1]}\{q,t=0\,|p\} =
\frac{q^{\alpha_{m,n}}}{ ||L_{Y_{m,n}}(q^{-1})||^2}
L_{Y_{m,n}}\left\{q^{-1} \Big|-\frac{1}{1-q^k}p_k\right\}
}
\label{Pt0}
\ee
Formula (\ref{Pq0}) is a trivial corollary of the fact that in
(\ref{HLexpan_0}) $h_Q \sim q^{\nu_Q}$ \cite{MMSS}.
Since $\nu_Q\geq 0$, only a single term with $\nu_Q=0$,
i.e. with $Q=[m]$ survives in the sum (\ref{chisup}) for $q=0$.
Notably,
since $h_Q=1$ for $l(Q)=1$,
this argument implies that the r.h.s. of eq.(\ref{Pq0}) is completely independent of $n$.

In  (\ref{Pt0})
\be
Y_{m,n} = \left[ \underbrace{ M + 1, \ldots, M + 1 }_{\sigma} ,
\underbrace{ M, \ldots, M }_{n - \sigma} \right],
\ee
where $M = \lfloor m/n \rfloor$, and $\sigma = m \mod n$,
so that $(M+1)\sigma + M(n-\sigma) = Mn + \sigma = m$.

Eq.(\ref{Pt0}) is related to (\ref{Pq0}) via the symmetry\footnote{
Note that the symmetry between $t$ and $q$ in the MacDonald polynomials is not quite
transparent: in fact,
$$
M_{Q}\{q,t|p_k\} = {(-1)^{|Q|}\over ||M_Q||^2} M_{Q'}\left\{t,q \left| - \dfrac{(1-t^k)}{(1-q^k)} p_k \right\}\right.
$$
This symmetry, together with the corresponding property of the MacDonald expansion coefficients
of the superpolynomials
$$
C^{Q}_{[1]}(q,t) = - q^{m-1} (qt)^{(m - 1)(r - 1)/2} (-1)^{|Q|} \dfrac{(1-q)}{(1-t)}
\frac{1}{||M_Q||^2} \, C^{Q'}_{[1]}(t^{-1},q^{-1})
$$
implies eq.(\ref{Symmet}).
}
\be
\bar P^{m,n}_{[1]}\{q,t|p_k\} =
- q^{m-1} (qt)^{(m - 1)(r - 1)/2}
 \ \dfrac{1-q}{1-t} \ \bar P^{m,n}_{[1]}\left\{t^{-1},q^{-1}\Big|-\frac{1-t^k}{1-q^k}p_k\right\},
\label{Symmet}
\ee
(it is generalization of the symmetry
$(q,t,A) \longleftrightarrow (t^{-1},q^{-1},A^{-1})$
at the topological locus (\ref{tolo}), introduced in \cite{DMMSS}).
The fact that the diagram $Y_{m,r}$ with $r$ lines appears in
(\ref{Pt0})
instead of $[1^m]$
is related to the restriction $l(Q)\leq r$ in (\ref{HLexpan_0}).

A simple example with $(m,n)=(3,2)$ illustrates, how (\ref{Pt0}) works:\footnote{
An even simpler example is $(m,n) = (2,n)$:
$$
\bar P^{2,n}_{[1]}\{q,t|p\}\ \stackrel{\cite{DMMSS}}{=}\
M_{[2]}\{q,t|p\}  - \frac{1-t^2}{1-qt}\sqrt{\frac{q}{t}}(qt)^{n/2}
M_{[11]}\{q,t|p\} =
$$
$$
= \frac{\Big((1+t)(1-q)+(1-t^2)\,q^{\frac{n+1}{2}}t^{\frac{n-1}{2}}\Big)p_2
+ \Big((1-t)(1+q)-(1-t^2)\,q^{\frac{n+1}{2}}t^{\frac{n-1}{2}}\Big)p_1^2}{2(1-qt)}
$$
The three particular cases of interest for us are:
$$q=0: \ \ \
\bar P^{2,n}_{[1]}\{q=0,t|p\} = \frac{1+t}{2}p_2 + \frac{1-t}{2}p_1^2 = L_{[2]}\{t|p\}
\ \ \ \ \ {\rm no}\ n\ {\rm dependence}
$$
$$ q=t:\ \ \
\bar P^{2,n}_{[1]}\{t,t|p\} = H^{2,n}_{[1]}\{t|p\} = \frac{1+t^n}{2}p_2 +
\frac{1-t^n}{2}p_1^2 = L_{[2]}\{t^n|p\}
$$
in accordance with eqs.(\ref{Pq0}) and (\ref{main_0}) respectively,
and
$$
t=0:\ \ \ \
\bar P^{2,n}_{[1]}\{q,t=0\,|p\} = \frac{1-q + q\delta_{n,1}}{2}p_2 + \frac{1+q-q\delta_{n,1}}{2}p_1^2
= \left\{\begin{array}{ccccc}
n=1: & & \chi_{[2]}\{p\} & = & -q(1-q)L_{[2]}\left\{\frac{1}{q}\Big|-\frac{p_k}{1-q^k}\right\}\\
n\geq 3: & & \frac{1-q}{2}p_2 + \frac{1+q}{2}p_1^2
& = & (1-q)(1-q^2)L_{[11]}\left\{\frac{1}{q}\Big|-\frac{p_k}{1-q^k}\right\}
\end{array}\right.
$$
Note that this time in eq.(\ref{Pt0})
the diagram $Y_{2/n}$ is different for $n=1$ and $n\geq 3$:
$Y_{2,1} = [2]$, while for $n\geq 3$ $M=0,\ \sigma=2$ and
$Y_{2,n} = [1,1,\underbrace{0,\ldots,0}_{n-2}] = [1,1]$.
Up to some powers of $q$
the normalization factors are $||L_{[2]}(q^{-1})||^{-2} = 1-q^{-1} = -q^{-1}(1-q)$
and $||L_{[11]}(q^{-1})||^{-2} = (1-q^{-1})^2(1+q^{-1}) = q^{-3}(1-q)(1-q^2)$
}
$$
\bar P_{[1]}^{3,2}\{t,q|p\} \stackrel{\cite{DMMSS}}{=}
M_{[3]}\{t,q|p\} + \frac{q^2(1-t)(1+t+qt)}{1-q^2t}M_{[2,1]}\{t,q|p\} +
\frac{q^3t(1-t)(1-t^2)(1+t+t^2)}{(1-qt)(1-q^2t)}M_{[1,1,1]}\{t,q|p\} =
$$
$$
= \frac{(1+t+t^2)(1-q+qt)}{3}p_3 + \frac{(1+qt)(1-t^2)}{2}p_2p_1 +
\frac{(1-t)^2(1+2q+qt)}{6}p_1^3 =
$$
\be
= L_{[3]}\{t|p\} + q(1-t)L_{[2,1]}\{t|p\}
\ee
Clearly, for $q=0$ we get
\be
\bar P^{3,2}_{[1]}\{q=0,t |p\} = L_{[3]}\{t|p\}
\ee
in accordance with (\ref{Pq0}), and for $q=t$
\be
\bar P^{3,2}_{[1]}\{t,t |p\} = \bar H^{3,2}_{[1]}\{t |p\} = L_{[3]}\{t|p\} + t(1-t)L_{[2]}\{t|p\}
= \nn \\
= \frac{1+t^2+t^4}{3}p_3 + \frac{1-t^4}{2}p_2p_1 + \frac{(1-t^2)^2}{6}p_1^3
= L_{[3]}\{t^2|p\}
\ee
in accordance with (\ref{main_0}).
For $t=0$ we reproduce (\ref{Pt0}) with $Y_{3,2} = [2,1]$, but calculation is a little more involved:
\be
\bar P^{3,2}_{[1]}\{q,t=0\, |p\} = \frac{1-q}{3}p_3 + \frac{1}{2}p_2p_1 + \frac{1+2q}{6}p_1^3
= q^2(1-q)^2 L_{[21]}\left\{\frac{1}{q}\Big|-\frac{1}{1-q^k}p_k\right\} =
\nn \\ =
q^2(1-q)^2 \left\{\frac{1+q^{-1}+q^{-2}}{3}\frac{p_3}{1-q^3} +
\frac{q^{-1}(1+q^{-1})}{2}\frac{p_2}{1-q^2}\frac{p_1}{1-q} -
\frac{(1-q^{-1})(2+q^{-1})}{6}\frac{p_1^3}{(1-q)^3}\right\}
\ee

In the HOMFLY case all the knots $[m,n]$ with the given $m$ are obtained
from $[m,r]$ by a simple evolution rule $t \longrightarrow t^{n/r}$.
In the above example
\be
\bar H_{[1]}^{3,n}\{t|p\} = \bar P_{[1]}^{3,n}\{t,t|p\} =
\frac{1+t^n+t^{2n}}{3}p_3 + \frac{1-t^{2n}}{2}p_2p_1 + \frac{(1-t^n)^2}{6}p_1^3
= L_{[3]}\{t^n|p\}
\ee
in particular,
$\bar P_{[1]}^{3,1}\{t,t|p\} = L_{[3]}\{t|p\}$
(and coincides with $\bar P_{[1]}^{3,2}\{q=0,t|p\}$).
However, for $q\neq t$ the evolution is not so simple:
one should decompose $P_{[1]}^{m,r}$ as a function of $\{p\}$
into a combination of MacDonald polynomials,
then the coefficient in front of $M_Q\{p\}$ is
multiplied by $\left(t^{\nu_Q}q^{-\nu_{Q'}}\right)^{\frac{n}{m}}$.
Since $M_Q$ themselves depend on $t$ and $q$,
this evolution is not reduced to any change of $q$ and $t$ parameters,
like it was in the HOMFLY case.
Still (\ref{Pq0}) and (\ref{Pt0}) hold for the entire series of
torus knots
introduced in \cite{DMMSS}, with all $n = mk+r$.

\section{Conclusion}

Character decompositions, like (\ref{chi}), (\ref{HLexpan_0}) or (\ref{main_0}),
are {\it much better} than any other approach to knot/braid polynomials,
because they provide an extension to all time-variables from the
"topological" locus (\ref{tolo}).

In \cite{MMSS} we demonstrated that these character expansions
(for the HOMFLY or superpolynomials, for ordinary or extended ones)
can be significantly simplified in the special bases.
Namely, we considered decompositions into the \HL characters.
In this letter we further elaborated on this subject
on the solid ground of the HOMFLY polynomials.

\bigskip

Our main conclusions are:
\begin{itemize}
\item
The \HL polynomials indeed provide an adequate expansion basis.
\item
However, even more natural than decomposition (\ref{HLexpan_0}),
which was considered in \cite{MMSS}, is an alternative one (\ref{main_0}),
where the quantum parameter $t^n$ of the \HL polynomial depends on $n$
(i.e. on the knot). For the torus knots this provides an explicit separation
of $m$ and $n$ variables (the symmetry between $m$ and $n$ is restored
only at the topological locus (\ref{tolo}).
\item
The colored HOMFLY polynomials (for the torus links/knots) are just
the \HL characters only for the fundamental representations $R=[1^s]$.
For other representations one still has only {\it decompositions}.
\item
Superpolynomials with $R=[1]$ reduce to a single Hall-Littlewood polynomial
not only in the HOMFLY case $q=t$, but also in the two other special cases
$q=0$ and $t=0$.
\end{itemize}

\bigskip

Of course, by themselves these \HL decompositions of torus HOMFLY
polynomials are not so much better than the naive character decomposition
(\ref{chi}).
However, there can be a considerable difference when one moves in
any of the two directions:
from the HOMFLY to superpolynomials, and from torus to generic knots.
For this purpose the knowledge of the relevant structures
is extremely important, and it looks like
that this hidden representation theory structure is much better
captured by the \HL
decompositions
than by the naive Schur and MacDonald ones.


Like in \cite{MMSS}, in this letter, we only report {\it facts}
and briefly discuss their immediate implications.
Motivations and proofs will be given elsewhere.

\section*{Acknowledgements}

Our work is partly supported by Ministry of Education and Science of
the Russian Federation under contract 14.740.11.0081, by NSh-3349.2012.2,
by RFBR grant 10-01-00536 and
by joint grants 11-02-90453-Ukr, 12-02-91000-ANF,
12-02-92108-Yaf-a, 11-01-92612-Royal Society.

\end{document}